\documentclass[a4paper,11pt,onecolumn]{article}
\usepackage{latexsym}
\usepackage{graphicx}

\title{\bf On the evolution of the magnetic field of Ap star $\alpha^2$ CVn }
\author{V.D. Bychkov$^{1}$, L.V. Bychkova$^{1}$, J. Madej$^{2}$, G.P. Topilskaya$^{3}$ \\
$^{1}$ Special Astrophysical Observatory, Russian Academy of Sciences; email: vbych@sao.ru\\
$^{2}$ Warsaw University Observatory, Poland; email: jm@astrouw.edu.pl   \\
$^{3}$ North-Caucasian Federal University, Stavropol, Russia; email: gtop@mail.ru  }

\date{    }
\begin{document}

\begin{titlepage}
\maketitle

\begin{abstract}
New high-precision measurements of the longitudinal magnetic field of
Ap stars suggest the existence of secular intrinsic variations of the 
global magnetic field in some stars. We argue that such changes are
apparent in the Ap star $\alpha^2$ CVn in the time scale of $\sim$ 10
years, which results from the analysis of literature data. Therefore, 
such an observation implies, that the rate of magnetic field evolution
of Ap stars is much higher than was previously thought.
\end{abstract}

  \begin{flushright}
    \vspace{2cm}
    {\it submitted to }  \\[2mm]
    {\rm Proc. of the IV Annual Int. Conf. } \\
    {\rm North-Caucasian Federal University} \\
    {\rm STAVROPOL, (Russia) }  \\
    {\rm 20-22 April 2016 }  \\[15mm]
  \end{flushright} 
\par\vspace{3mm}

\end{titlepage}

\section{Introduction}

The study of evolution of the global magnetic fields of Ap stars is
a very important recent research subject.
Theoretical studies of the evolution of these objects were carried out
by a number of authors and were extensively summarized by Moss (1990).
According to his results, magnetic fields of stars also must evolve in 
the time scale of a general evolution of stars, i.e. in time scale of about 
$2.5\times 10^7 - 1.31\times 10^8$ years.
Landstreet et al. (2007) presented the most accurate estimates of the rate
of magnetic field evolution of Ap stars obtained on the basis of
observational data.

Estimates of the characteristic time of the magnetic field evolution
yield $2-3\times 10^7$ years for massive A stars of mass $M$ higher than
3 solar masses, $M> 3M_\odot$. For stars with masses lower than 3 solar 
masses, $M< 3M_\odot$, the rate of evolution is of the order of $10^8$ 
years. These numbers basically refer to stars which have the global magnetic
field of a simple dipole structure and are fully
consistent with theoretical estimates by Kochukhov and Bagnulo (2006).

According to theoretical studies by Krause and Raedler (1980), if the 
global magnetic field has a more complex structure of the quadrupole, 
octupole, etc., then its evolution proceeds much more rapidly than in 
the case of a simple dipole.

\section{New observational data}

Currently, accuracy of determination of the longitudinal magnetic field 
in stars has
substantially improved due to the use of modern light detectors and 
new methods of processing of observational data. New techniques revealed 
a number of subtle effects in the run of the longitudinal magnetic field
variations, which previously were unavailable for research. 

Two series of projects were recently carried out on exact measurements
of $B_e$ in stars during past two decades, which enabled us for the 
construction of the magnetic phase curves of some Ap stars. 
Results of the first high-precision measurements of the longitudinal 
magnetic fields $B_e$ using the Least Squares Deconvolution (LSD) method
(Donati et al. 1997) and WLSD (Wade et al. 2000a), obtained at the end 
of the XXth century, were published in Wade et al. (2000b). In the latter
study, series of $B_e$ measurements were obtained for 14 Ap stars and for 
11 of stars in that set observations cover all phases of the rotational
period. Silvester et al. (2012) also published $B_e$ series obtained by
analogous methods for 7 Ap stars. 

The latter measurements were carried out in
years 2006 - 2010 and for three stars they satisfactorily cover 
all phases of the rotational period. In both quoted papers
the average $B_e$ measurement error corresponding to $1\sigma$ is lower 
than 50 G. There are 6 stars studied in both sets of $B_e$ measurements for which 
authors presented data on two high-precision phase curves separated by several
years of observations and obtained using the same methodology. Therefore,
these phase curves can be directly compared in order to search for possible
changes in the magnetic phase curves of CP stars for 10 years.

\subsection{$\alpha^2$ CVn}

Significant changes of the magnetic phase curve showed up in the well-studied 
magnetic Ap star $\alpha^2$ CVn (SiCrEuHg type). Periodic variability of
the longitudinal (effective) magnetic field $B_e$ with the rotational phase
has a complex double-wave curve which corresponds to the structure of a quadrupole
magnetic field (at the first approximation). Results of the first high-precision
measurements using the LSD method were first publishd by Donati et al. (1997)
and Wade et al. (2000b). Authors of the latter paper presented a series of
18 measurements distributed over 700 days (1.9 years) around the average
date JD 2550848.56, see open circles in Fig. 1.
It was found that the average accuracy of
$B_e$ measurements equals 27.8 G.

\begin{figure}[t!]
\includegraphics[width=12cm]{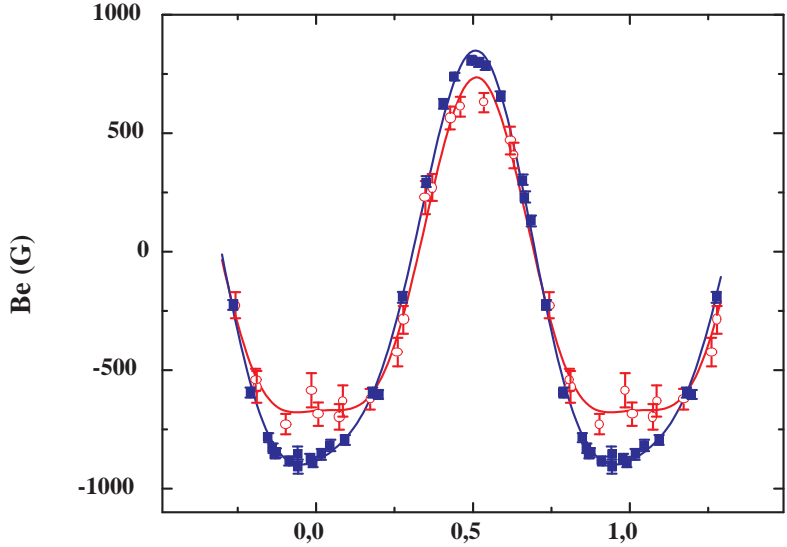}
\caption{Magnetic phase curves of $\alpha^2$ CVn. Red line denotes phase
   curve derived from $B_e$ points in Wade et al. (2000), whereas black 
   line was derived from Silvester et al. (2012). }
\end{figure}

The second set of high-precision measurements was obtained by the same
methods and was published in Silvester et al. (2012). Their results form
a series of 27 measurements obtained over 1295 days (3.5 years) with the
average date JD 2554722.51. These estimates are plotted in Fig. 1 as 
filled circles with the corresponding magnetic phase curve. Phase curve
and its amplitude was found with the average accuracy of 9.4 G (solid line).

Rotational phases of $\alpha^2$ CVn in Fig. 1 were determined using
the ephemeris by Farnsworth (1932), $JD(EuII max)=2419869.720
+5.46939\, E$.

Time interval between the centers of these two sets equals 3874 days 
(10.6 years). Figure 1 clearly shows that during the period of about 
10 years phase curve markedly changed. Calculated difference between 
both phase curves is plotted as the function of the rotational phase 
in Figure 2, which also shows the $1\sigma$ uncertainty of the amplitude
of the residual phase curves equal to 29.4 G.

\begin{figure}[t!]
\includegraphics[width=12cm]{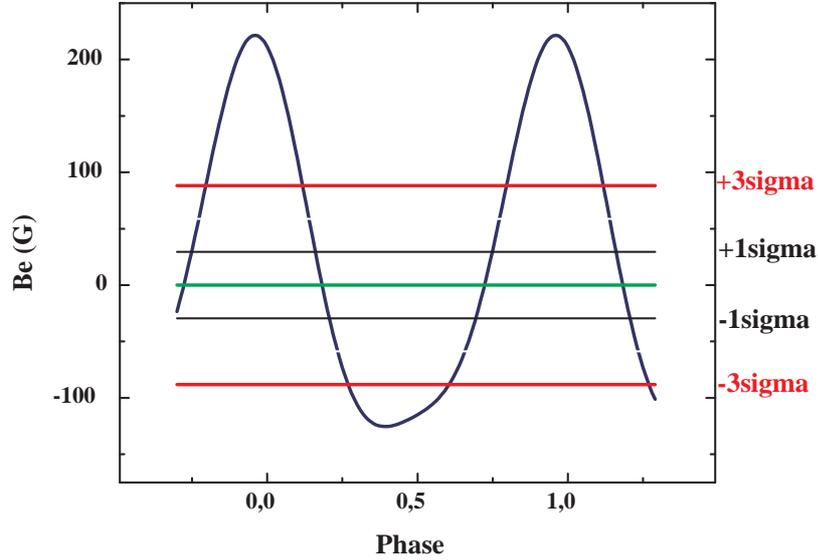}
\caption{Difference between both phase curves of $\alpha^2$ CVn,
   $B_e [2000] - B_e [2012]$, is given by the solid line. The amplitude
   of the residual phase curve strongly exceeds the $3\sigma$ uncertainty
   range of the magnetic phase curve determination.  }
\end{figure}

The largest difference of 222 G occurs in the rotational phase 0.96,
which equals 7.5 sigma. In phase of 0.39 the difference between both 
phase curves equals -125 G, or 4.3 sigma. Therefore, it is a very
significant difference. It is very likely that this is a real change
of the magnetic phase curve occuring during about 10 years,
Therefore, we note that the rate of evolution of the magnetic field in this
star by about 5 - 6 orders of magnitude faster than was previously thought.

It should be noted that, according to Krause and Raedler (1980), evolution
of the global magnetic fields of stars with a complicated structure (quadrupole,
octupole, etc.) should take place significantly faster than in stars with
a simple dipole structure of the magnetic field. But even in such a  case
it turns out that the magnetic field of $\alpha^2$ CVn actually evolves 
much faster (4 - 5 orders of magnitude) than was predicted before 
(Krause and Raedler 1980).

\section{Discussion}

Naturally the following question arises - how realistic is this effect? 
To confirm that this is a real feature we demonstrate lack of such phase 
curves differences in some other Ap stars in which the longitudinal magnetic
field was measured by the same instrument and methods, published by the same
authors in the same papers. We have not found such significant changes
of the magnetic phase curves for the other CP stars from the list. 

As an example, consider magnetic Ap star HD62140.
In the paper by Wade et al. (2000a), 14 high-precision measurements of
the longitudinal magnetic field $B_e$ were obtained during 705 days of observations
(1.9 years) centered on JD 24550858.0. Fig. 3 presents two magnetic phase
curves, again the second curve was derived from Silvester et al. (2012).
The latter paper presented 19 measerements obtained during 1171 days
(3.2 years) about JD 24554666.3.

\begin{figure}[ht!]
\includegraphics[width=12cm]{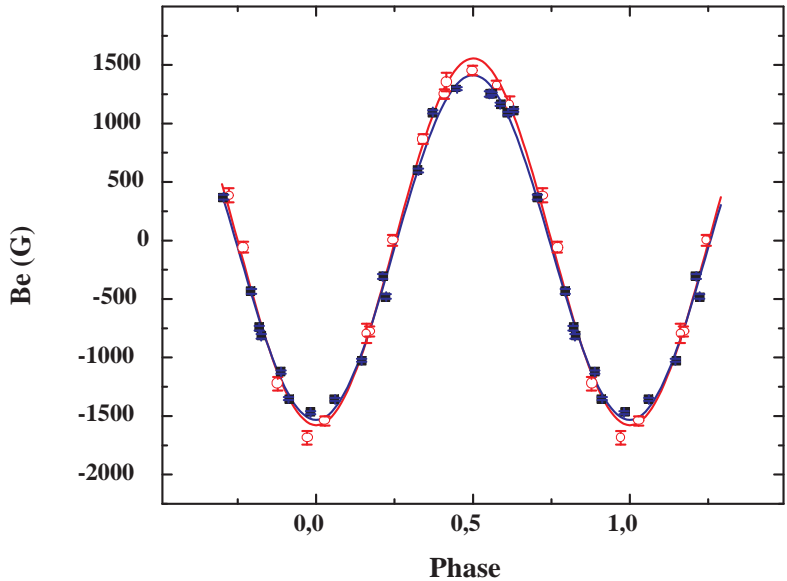}
\caption{Phase curves of HD62140. Open circles - $B_e$ measurements by
    Wade et al. (2000a), filled circles - Silvester et al. (2012).  }
\end{figure}

Time span between the centers of both sets of $B_e$ measurements equals 
3808.3 days (10.4 years). Duration of sets, number of $B_e$ points and the
time interval between sets is very close to that obtained for $\alpha^2$ CVn.
At the rotational phase $\psi=0.01$ the difference between phase curves amounts
to $1.5 1\sigma$, for $\psi = 0.51$ to $4.9 1\sigma$.

For HD62140 long-term changes of the phase curve are much lower and apparently
show the opposite sign. I.e. the amplitude value estimates for $\alpha^2$ CVn increased
as compared with the first set (phase change and the shape of the curve), while for the opposite HD62140 slightly
decreased in high magnetic field.
Similar results were obtained for other stars have been investigated in these studies.

Fig. 4 shows results of another set of high-precision of $B_e$ measurements of
HD71866, derived from the same two papers using the same methods, instrument
and similar dates of observation.
 
\begin{figure}[ht!]
\includegraphics[width=12cm]{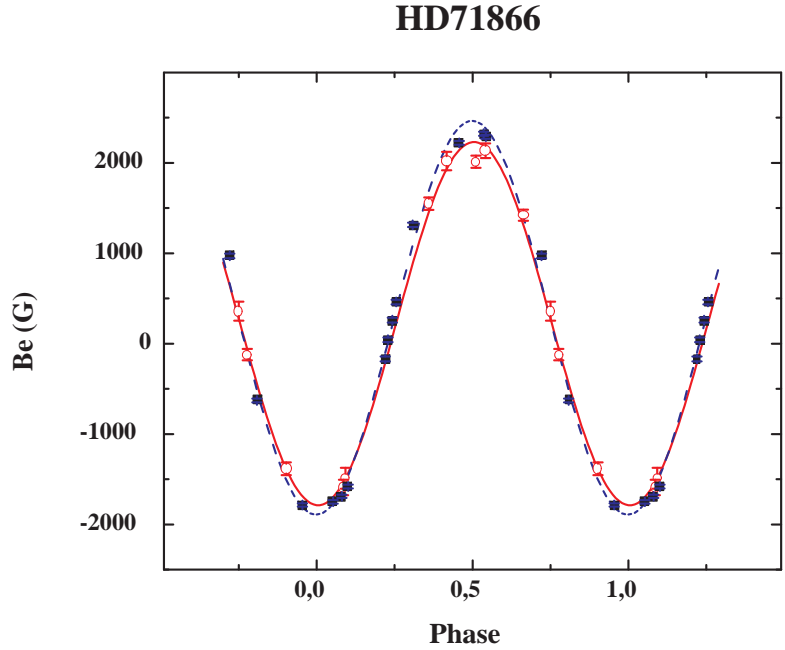}
\caption{Phase curves of HD71866. Open circles - Wade et al. (2000a), 
    filled circles - Silvester et al. (2012). }
\end{figure}

\section{ Conclusions}

We noted the probable existence of secular changes of the global
magnetic field in Ap star $\alpha^2$ CVn exceeding the estimated
$3\sigma$ uncertainty level. Our observation was dervived from two 
published papers which presented series of the longitudinal magnetic
field $B_e$ of a small group of Ap stars obtained by the precise
Least Squares Deconvolution method. We compared here $B_e$ rotational
phase curves of $\alpha^2$ CVn separated by a time span of $\approx$ 10 years.

At present it is still too early to draw definite conclusions on the
prompt secular evolution of the magnetic field of $\alpha^2$ CVn. 
That problem requires obtainin of more high-precision observational 
data. Actually we prepare new instrument at SAO in order to verify the
complex and subtle problem of secular variations of the magnetic field in Ap stars
(Valyavin et al. 2014). If the change of the magnetic phase curve of
$\alpha^2$ CVn is confirmed, it will radically change the whole idea of the evolution of global
magnetic fields Ap stars.

\section{Acknowledgements}

Research by V.D. Bychkov was supported by the Russian Scientific
Foundation grant N14-50-00043.

\end{document}